\documentclass[twocolumn,notitlepage,10pt,pre,superscriptaddress,showpacs]{revtex4-1}
\usepackage[english]{babel}
\usepackage{amsmath}
\usepackage{amssymb}
\usepackage{bm}
\usepackage{graphicx}
\usepackage{mathptmx}
\usepackage{color}

\begin{document}

\title{Exchange effects in plasmas: the case of low-frequency dynamics}
\author{J. Zamanian}
\email[E-mail address: ]{jens.zamanian@physics.umu.se}
\affiliation{Department of Physics, Ume{\aa} University, SE-901 87 Ume{\aa}, Sweden}
\author{M. Marklund}
\email[E-mail address: ]{mattias.marklund@chalmers.se}
\affiliation{Department of Physics, Ume{\aa} University, SE-901 87 Ume{\aa}, Sweden}
\affiliation{Department of Applied Physics, Division for Condensed Matter Theory, Chalmers University of Technology, SE-412 96 G\"oteborg, Sweden}
\author{G. Brodin}
\email[E-mail address: ]{gert.brodin@physics.umu.se}
\affiliation{Department of Physics, Ume{\aa} University, SE-901 87 Ume{\aa}, Sweden}

\begin{abstract}
Recently, there has been a surge in the interest of non-equilibrium
collective quantum models, where particle dispersion and spin are examples
of effects taken into account. Here, we derive a kinetic plasma model
containing \textcolor{black}{fermion} exchange effects. 
\textcolor{black}{Exchange interactions are of great importance in many
systems}, and have no classical analogy. Our model therefore constitute a
possible probe of collective quantum phenomena 
\textcolor{black}{in new
regimes}. As an example, we consider the influence of exchange effect on low
frequency dynamics, in particular ion acoustic waves. Comparisons to related
computational techniques are given and the differences are highlighted.
Furthermore, we discuss the applicability of our model, %
\textcolor{black}{its limitations} and possible extensions.
\end{abstract}

\pacs{52.25.-b, 52.25.Dg}
\maketitle

\affiliation{Institut de Physique et Chimie des Mat\'eriaux de Strasbourg,
23 Rue du Loess, BP 43, F--67034 Strasbourg, France}

\affiliation{Applied Physics, Chalmers University of Technology, SE--412 96
G\"oteborg, Sweden} 
\affiliation{Department of Physics, Ume{\aa} University, SE--901 87
Ume{\aa}, Sweden}

\affiliation{Department of Physics, Ume{\aa} University, SE--901 87
Ume{\aa}, Sweden}

%Generally about quantum plasmas
%Some regimes
%Exchange effects typically small... 
%Have been studied in other areas such as... 
%Dft, kinetic, greens functions 
%In this paper... 

\section{Introduction}

Quantum plasma physics is currently a field of intense study. One reason for
this is the potential applications in, for example, laser produced plasmas \cite{Mourou-etal,Marklund-Shukla,DiPiazza-etal,Glenzer-Redmer},
ultra small electronic devices, and dense astrophysical systems \cite%
{manfredi2006,Haas-book,Shukla-Eliasson-RMP}. Different aspects of quantum
plasmas have been studied such as quantum dispersion and Fermi pressure \cite%
{manfredi2006,Haas-book,Shukla-Eliasson-RMP}, the magnetic dipole force and
the spin dynamics \cite{burt62, cowley86,brodin08-1, morandi09, zamanian07,
shukla10, brodin10, zamanian10-1,stefan11,brodin08}, quantum relativistic
effects and nonlinear dynamics \cite{shukla10, shaikh07, brodin10,
stefan11,manfredi2005}. Typically, quantum effects are important for systems
with high density and low temperature. 
\textcolor{black}{This said, it is
important to distinguish between quantum effects related to thermodynamic
equilibrium properties and dynamical properties of the system. } Exchange
effects due to particle statistics have been successfully included in the
density functional theory (DFT) \cite{hohnenberg64,khon65}. Applications of
DFT include for example ground state properties of atoms and equilibrium
properties of many-particle systems \cite{hohnenberg64,khon65}. The effects
of exchange on dynamics have also been studied in the setting of kinetic
theory \cite%
{vonroos60,vonroos61,burt62,klimontovich74,ma11,bonitz,brosnens84,nacht83},
as well as in studying, e.g., the thermodynamic properties of plasmas \cite%
{klimontovich74,bonitz}. It has also been studied using fluid theory \cite%
{ma11}. Furthermore, many papers deal with how quantum mechanics affects the
low-frequency long-scale dynamics, as for example quantum ion acoustic waves 
\cite{haas03, misra06, mahmood08, mushtaq09, gill10}.

In Section \ref{model} we derive the Wigner equation for electrons within
the Hartree-Fock approximation. We simplify the equation by assuming that
the plasma is not spin-polarized and by focusing on length scales much
longer than the thermal de Broglie wave length. In Section \ref{ionacoustic}
we consider the impact on ion-acoustic waves by treating the exchange
effects perturbatively within the linear approximation and finally in
Section \ref{discussion} we discuss our result.

%%%%%%%%%%%%%%%%%%%%%%%%%%%%%
%Model 
%%%%%%%%%%%%%%%%%%%%%%%%%%%%%

\section{Model}

\label{model} We here consider a completely ionized electron-ion plasma with
the particles interacting through a mean-field scalar potential. Quantum
effects for the ions will be completely neglected, while for the electrons
we will take into account a dynamic correction due to the Pauli exclusion
principle. Also, we will not consider effects due to the self-energy and
particle correlations \cite{bonitz}. We here give an outline of the
derivation of a kinetic theory with exchange effects.

The state of the $N$-electrons is described by the density operator $\rho
_{1\dots N}$ (see for example Ref.\ \cite{bonitz}), and the dynamics is
given by the von Neumann equation with the Hamiltonian 
\begin{equation}
\hat{H}_{1\dots N}=\sum_{i=1}^{N}\frac{\hat{p}_{i}^{2}}{2m_{e}}+\frac{e^{2}}{%
4\pi \epsilon _{0}}\sum_{i<j}\frac{1}{|\hat{\mathbf{x}}_{i}-\hat{\mathbf{x}}%
_{j}|}+e\sum_{i=1}^{N}\varphi (\hat{\mathbf{x}}_{i}).
\end{equation}%
Here $m_{e}$ is the electron mass, $e$ is the electron charge ($e<0$) and $%
\epsilon _{0}$ is the permittivity of vacuum. The last term accounts for the
interaction with the electric potential created by the ions. We now
introduce the reduced density operators according to 
\begin{equation}
\hat{\rho}_{1\dots i}=n^{i}\mathrm{Tr}_{i+1\dots N}\hat{\rho}_{1\dots N}\hat{%
\Lambda}_{1\dots i},
\end{equation}%
where $\mathrm{Tr}_{i+1\dots N}$ denotes the trace over particles $i+1$ to $%
N $ (i.e. integrating over the position degree of freedom and summing over
the spins), $n$ is the mean density and $\hat{\Lambda}_{1\dots i}$ is the
antisymmetrization operator that takes an $i$-particle state and makes it
completely antisymmetric \cite{db79}. We will only need to know that $\hat{%
\Lambda}_{12}=1-\hat{P}_{12}$ where $\hat{P}_{12}$ interchanges particle 1
and 2, i.e. $\hat{P}_{12}\psi (\mathbf{x}_{1},\mathbf{x}_{2})=\psi (\mathbf{x%
}_{2},\mathbf{x}_{1})$ (see, e.g., Ref.\ \cite{bonitz} for further details).
The evolution for the one-particle density operator is given by 
\begin{equation}
i\hbar \partial _{t}\hat{\rho}_{1}=[\hat{h}_{1},\hat{\rho}_{1}]+n\mathrm{Tr}%
_{2}[\hat{V}_{12},\hat{\rho}_{12}\hat{\Lambda}_{12}],  \label{rho1}
\end{equation}%
where $\hat{h}_{1}=\hat{p}^{2}/(2m_{e})$ and $\hat{V}_{12}=V(\hat{\mathbf{x}}%
_{1}-\hat{\mathbf{x}}_{2})=e^{2}/(4\pi \epsilon _{0}|\hat{\mathbf{x}}_{1}-%
\hat{\mathbf{x}}_{2}|)$ and $\hat{\rho}_{12}$ is the two-particle density
operator. The effects of two-particle correlations $\hat{g}_{12}$ can be
separated out of the two-particle density operator by writing it in the form 
\begin{equation}
\hat{\rho}_{12}=\hat{\rho}_{1}\hat{\rho}_{2}+\hat{g}_{12},
\label{correlations}
\end{equation}%
see e.g. Ref.\ \cite{wang85}. We are interested in the collisionless limit
where a mean-field approximation will suffice. This approximation is
obtained by neglecting the correlation $\hat{g}_{12}$. Utilizing this in (2)
we obtain 
\begin{equation}
i\hbar \partial _{t}\hat{\rho}_{1}=[\hat{h}_{1},\hat{\rho}_{1}]+[\bar{V}_{1},%
\hat{\rho}_{1}],  \label{rhomf}
\end{equation}%
where $\bar{V}_{1}=\mathrm{Tr}_{2}\hat{V}_{12}\hat{\rho}_{2}\hat{\Lambda}%
_{12}$, is the Hartree-Fock potential operator. This is a closed system for
the one-particle density operator.

To obtain a connection to the classical kinetic theory we utilize the Wigner
representation \cite{wigner32} of this equation. Using the complete set of
states $\left| \mathbf{x}, \alpha \right>$, where $\mathbf{x}$ is the
position and $\alpha = 1,2$ is the spin along the axis of quantization, this
representation is obtained as 
\begin{eqnarray}
f(\mathbf{x}, \mathbf{p}, \alpha, \beta ) &=& \frac{n}{(2\pi\hbar)^{3} }
\int d^3\! y \,\, e^{i \mathbf{y }\cdot \mathbf{p }/ \hbar} \rho \left( 
\mathbf{x }+ \frac{\mathbf{y}}{2} , \alpha ; \mathbf{x }- \frac{\mathbf{y}}{2%
} , \beta \right),  \notag \\
\end{eqnarray}
where $\rho(\mathbf{x}, \alpha ; \mathbf{y }, \beta) = \left< \mathbf{x},
\alpha \right| \hat \rho_1 \left| \mathbf{y}, \beta \right>$ is the density
matrix. Writing Eq.\ \eqref{rhomf} first in the position representation and
Wigner transforming the result we obtain 
\begin{widetext}
\begin{align}
	& \partial_t  f ( \mathbf x , \mathbf p , \alpha, \beta) 
	+ \frac{1}{m} \mathbf p \cdot \nabla_x f(\mathbf x , \mathbf p, \alpha, \beta)    
	+ \frac{i e}{\hbar} \int  \frac{ d^3\! y \, d^3\! p'}{(2 \pi \hbar)^3} \, 
	e^{i \mathbf y \cdot ( \mathbf p - \mathbf p')/ \hbar} 
	\left[ \phi \left( \mathbf x + \frac{\mathbf y}{2} \right) 
		- \phi \left( \mathbf x - \frac{\mathbf y}{2} \right) \right] 
	f( \mathbf x, \mathbf p' , \alpha,  \beta) 
	\notag \\ 
	& \qquad  =
	  \frac{i}{\hbar(2\pi \hbar)^3} \sum_{\gamma=1}^2 \int d^3\! p' \, d^3\! p'' \, d^3\! y \, d^3\! r \,\,
	 e^{ i \mathbf p \cdot \mathbf y/ \hbar } 
	 e^{ - i \mathbf p' \cdot ( \mathbf x + \mathbf y /2 - \mathbf r)/\hbar} 
	 e^{ - i \mathbf p'' \cdot ( \mathbf r - \mathbf x + \mathbf y /2 ) / \hbar} 
	 \notag \\
	 & \qquad\qquad \times \left[ V \left( \mathbf x + \frac{\mathbf y}{2} - \mathbf r \right) 
	 - V \left( \mathbf x - \frac{\mathbf y}{2} - \mathbf r \right) \right] 
	 f \left( \frac{\mathbf x + \mathbf r}{2} + \frac{\mathbf y}{4} , \mathbf p' , \alpha, \gamma \right) 
	 f \left( \frac{\mathbf x + \mathbf r}{2} - \frac{\mathbf y}{4} , \mathbf p'' , \gamma, \beta \right) ,
	 \label{14} 
\end{align} 
\end{widetext}
where 
\begin{equation}
\phi (\mathbf{x}) = \frac{e n}{4 \pi \epsilon_0} \sum_{\gamma=1}^2 \int
d^3\! z \,\, \frac{\rho(\mathbf{z }, \gamma; \mathbf{z} ,\gamma )}{| \mathbf{%
x }- \mathbf{z }|} + \varphi(\mathbf{x}).
\end{equation}
is the total (mean-field and the ionic field) potential and 
\begin{equation}
V(\mathbf{x}) = \frac{e^2}{4\pi \epsilon_0 |\mathbf{x}| }
\end{equation}
is the Coulomb potential. The left hand side of Eq.\ \eqref{14} represents
the quantum Vlasov equation, while the right hand side is the correction due
to exchange effects. This term is nonlocal in phase-space and nonlinear in
the distribution function.

The matrix equation can be transformed into a scalar equation by taking the
spin transformation \cite{zamanian07} 
\begin{equation}
f(\mathbf{x}, \mathbf{p}, \mathbf{s}, t) = \frac{1}{4\pi} \sum_{\alpha,
\beta = 1}^2 \left[ \delta_{\alpha, \beta} + \mathbf{s }\cdot \bm %
\sigma_{\alpha, \beta} \right] f(\mathbf{x}, \mathbf{p}, \beta, \alpha) ,
\end{equation}
where $\mathbf{s}$ is a vector on the unit sphere. Applying this to Eq.\ %
\eqref{14} we obtain 
\begin{widetext}
\begin{align}
	& \partial_t  f ( \mathbf x , \mathbf p , \mathbf s) 
	+ \frac{1}{m} \mathbf p \cdot \nabla_x f(\mathbf x , \mathbf p, \mathbf s)    
	+ \frac{i e}{\hbar} \int  \frac{ d^3\! y \, d^3\! p'}{(2 \pi \hbar)^3} 
	e^{i \mathbf y \cdot ( \mathbf p - \mathbf p')/ \hbar} 
	\left[ \phi \left( \mathbf x + \frac{\mathbf y}{2} \right) 
		- \phi \left( \mathbf x - \frac{\mathbf y}{2} \right) \right] 
	f( \mathbf x, \mathbf p' , \mathbf s) 
	\notag \\ 
	& \qquad  =
	  \frac{i}{ \hbar} \int \frac{d^3\! p' \, d^3\! p'' \, d^3\! y \, d^3\! r}{(2\pi\hbar)^3} \,\,
	 e^{ i \mathbf p \cdot \mathbf y/ \hbar } 
	 e^{ - i \mathbf p' \cdot ( \mathbf x + \mathbf y /2 - \mathbf r)/\hbar} 
	 e^{ - i \mathbf p'' \cdot ( \mathbf r - \mathbf x + \mathbf y /2 ) / \hbar} 
%\notag \\ 
%	& \qquad \qquad \qquad
	\int \frac{d^2 s' d^2 s''}{8\pi} \left[ 1 +  9 \mathbf s' \cdot \mathbf s''+ 3 \mathbf s \cdot ( \mathbf s' + \mathbf s'') 
	+ 9 i \mathbf s \cdot \left( \mathbf s' \times \mathbf s'' \right)
	\right] 
	 \notag \\
	 & \qquad\qquad 
	  \times \left[ V \left( \mathbf x + \frac{\mathbf y}{2} - \mathbf r \right) 
	 - V \left( \mathbf x - \frac{\mathbf y}{2} - \mathbf r \right) \right] 
	 f \left( \frac{\mathbf x + \mathbf r}{2} + \frac{\mathbf y}{4} , \mathbf p'  ,\mathbf s' \right) 
	 f \left( \frac{\mathbf x + \mathbf r}{2} - \frac{\mathbf y}{4} , \mathbf p'' , \mathbf s'' \right) ,
	 \label{14-spin} 
\end{align} 
\end{widetext}
where in the last term we see the exchange interaction in the Wigner form.
The evolution equation \eqref{14} describes the evolution of the electrons
in the mean-field (Hartree-Fock) approximation for all scale lengths. We are
interested in the semiclassical limit where the potential $\phi$ and the
distribution function $f$ vary on a scale $L$ much larger than the de
Broglie scale length $\Lambda_{\mathrm{dB}}$ and would like to keep only the
lowest surviving correction in an expansion in $\Lambda_{\mathrm{dB}}/L$.
For the potential term the expansion is straightforward, see for example
Ref.\ \cite{manfredi2006}. For the exchange term we expand the potential and
the distribution function to second order in $\mathbf{y}$ (with the
assumptions that the characteristic length scale $L$ is much larger than the
thermal de Broglie wave length $\hbar /m v_T$, where $v_T$ is the thermal
speed). We then perform the $\mathbf{y}$-integration and one of the momentum
integrals.

Furthermore, we will for simplicity also assume that the distribution
function is independent of the spin, i.e. $f(\mathbf{x},\mathbf{p}, \mathbf{s%
},t) = f(\mathbf{x}, \mathbf{p},t) / (4 \pi)$. Integrating over the spin we
obtain 
\begin{widetext} 
\begin{align} 
	\partial_t f (\mathbf x, \mathbf p, t) 
	+ \frac{\mathbf p}{m} \cdot \nabla_x f(\mathbf x, \mathbf p, t) 
	+ e \mathbf E (\mathbf x,t) \cdot \nabla_p f(\mathbf x ,\mathbf p, t) 
	 = & \,
	 \frac{1}{2} \partial_p^i \int d^3\! r \, d^3\! p' \,\, e^{ - i \mathbf r \cdot \mathbf p' / \hbar} 
	[\partial_r^i V (\mathbf r)] 
	f \left( \mathbf x - \frac{\mathbf r}{2} , \mathbf p + \frac{\mathbf p'}{2}, t \right) 
	f \left( \mathbf x - \frac{\mathbf r}{2} , \mathbf p - \frac{\mathbf p'}{2}, t \right)  
\notag \\ & 
	- \frac{i \hbar}{8}  
	\partial_p^i \partial_p^j \cdot \int d^3\! r \, d^3\! p' \,\, e^{ - i \mathbf r \cdot \mathbf p' / \hbar} 
	[\partial_r^i V (\mathbf r) ]
\notag \\ & 
	\quad \quad \quad \quad \quad 
	\times
	\left[ f \left( \mathbf x - \frac{\mathbf r}{2} , \mathbf p - \frac{\mathbf p'}{2},  t \right) 
	\left( \overleftarrow \partial_x^j  - \overrightarrow \partial_x^j \right)
	f \left( \mathbf x - \frac{\mathbf r}{2} , \mathbf p + \frac{\mathbf p'}{2} , t \right)  \right]
\label{23}  
\end{align} 
\end{widetext} 
where $\partial_x^i \equiv \partial/\partial x_i$ and analogously for $%
\partial_p^i$ and an arrow above an operator indicates in which direction it
acts. We have also used the summation convention so that a sum over indices
occurring twice in a term is understood.

Eq. (\ref{23}) is one of the main results of this work. Unfortunately the
complexity of this equations limits the practical applicability to some
extent. The key advantage is that it is derived from first principles using
few assumptions beside the perturbative approach. Theories that can be
applied in the regime of stronger exchange effects by necessity use
assumptions and/or approximations that need validation. In some cases
justification can be done against experiments, but comparison against theory
based on first principles are also much needed. Such comparisons need to be
done in the regime of weak exchange effects, as it is only here calculations
based solely on first principles can be made. However, since Eq. (\ref{23})
can be used as a validation for other calculation schemes (like e.g. density
functional theory), it has relevance beyond the regime of weak exchange
effects. The final section of the manuscript will elaborate on these issues.

%%%%%%%%%%%%%%%%%%%%%%%%%%%%%%%%%%%%
%Ion-accoustic waves
%%%%%%%%%%%%%%%%%%%%%%%%%%%%%%%%%%%%

\section{Damping of ion-acoustic waves}

\label{ionacoustic} We now consider the effect of the exchange term on
electrostatic ion-acoustic waves in a plasma. We will use Eq.\ \eqref{23}
for the electrons and the classical Vlasov equation for ions. 
%\begin{widetext} 
%\begin{align} 
%	\partial_t f (\mathbf x, \mathbf p, t) & + \frac{1}{m} \mathbf p \cdot \nabla_x f(\mathbf x, \mathbf p, t) 
%	+ e \mathbf E (\mathbf x,t) \cdot \nabla_p f(\mathbf x ,\mathbf p ,t) 
%	\notag \\ =
%	& 
%	\frac{1}{2} \nabla_p \cdot  \int d^3\! r \, d^3\! p' \,\, e^{ - i \mathbf r \cdot \mathbf p' / \hbar} 
%	[\nabla_r V (\mathbf r)] 
%	f \left( \mathbf x - \frac{\mathbf r}{2} , \mathbf p + \frac{\mathbf p'}{2} \right) 
%	f \left( \mathbf x - \frac{\mathbf r}{2} , \mathbf p - \frac{\mathbf p'}{2} \right)  
%	\notag \\ &
%	- \frac{i \hbar}{8}  
%	\partial_p^i \partial_p^j \cdot \int d^3\! r \, d^3\! p' \,\, e^{ - i \mathbf r \cdot \mathbf p' / \hbar} 
%	[\partial_r^i V (\mathbf r) ]
%	\left[ f \left( \mathbf x - \frac{\mathbf r}{2} , \mathbf p + \frac{\mathbf p'}{2} \right) 
%	\partial_x^j f \left( \mathbf x - \frac{\mathbf r}{2} , \mathbf p - \frac{\mathbf p'}{2}
%	 \right)
%	 \right. 
%	 \notag \\ &
%	 \left. \quad \quad \quad \quad \quad \quad \quad \quad \quad \quad \quad \quad
%	 \quad \quad \quad \quad \quad \quad \quad \quad
%	 - f \left( \mathbf x - \frac{\mathbf r}{2} , \mathbf p - \frac{\mathbf p'}{2} \right) 
%	\partial_x^j f \left( \mathbf x - \frac{\mathbf r}{2} , \mathbf p + \frac{\mathbf p'}{2}
%	 \right)
%	  \right] . 
%\label{24}  
%\end{align} 
%\end{widetext} 
To obtain the dispersion relation we assume a longitudinal oscillation $%
f=f_{0}(p)+f_{1}(\mathbf{p})\exp (-i\omega t+ikz)$ and $\mathbf{E}=\hat{%
\mathbf{z}}E_{z}\exp (-i\omega t+ikz)$. We assume that the unperturbed
electron distribution function is given by a Maxwell-Boltzmann distribution 
\cite{maxwell-note} 
\begin{equation}
f_{0}(\mathbf{p}) = \frac{n}{( 2\pi m k_B T_e )^{3/2} } \exp \left( -\frac{%
p_{\perp}^{2}+p_{z}^{2}}{2 m k_{B} T_e } \right) .
\end{equation}%
Furthermore, assuming that the exchange terms are small correction to the
distribution function, we may calculate it by inserting the lowest order
solution for $f_{1}$, i.e. $f_1 = - i e E_z / (\omega - k p_z / m )
\partial_{p_z} f_0$, in the integrand. Introducing spatial spherical
coordinates, it is possible to solve all spatial integrals in Eq.\ \eqref{23}%
. Next, the integrand is expanded in terms of $\hbar $. The lowest order
term in the first integral vanishes due to symmetry and we keep only the two
first-order terms. In the second integral we already have an additional $%
\hbar $, meaning that we only retain the lowest order term. Finally it is
possible to solve the $p_{z}^{\prime }$ and $\varphi _{p}^{\prime }$
integrals. The remaining integrals are solved numerically and doing so gives
a solution for $f_{1}$ in the linear regime. Now, from the classical
dispersion relation we have 
\begin{equation}
\omega \approx (\omega_{pI}/\omega_{pe}) kv_{Te} \equiv \alpha kv_{Te} ,
\end{equation}
where $v_{Te} = \sqrt{k_B T_e/m_e}$ is the electron thermal velocity and $%
\omega_p$ denotes the plasma frequency. The dispersion relation is then
given by 
\begin{widetext}
\begin{align} 
	0 = & \:
	1 + \frac{\omega_{pe}^2}{k^2 v_{Te}^2} - \frac{\omega_{pI}^2}{\omega^2} 
	- \frac{\hbar^2 \omega_{pe}^4}{4\pi m^2 k^2 v_{Te}^6 } 
	\int  dv \frac{e^{ -v^2 } }{ ( \alpha - v)^2} \int du 
	\left[  \frac{v + u}{ \alpha -  ( v + u) }\right] 
	\left[ \left( u^2 - \frac{u}{\alpha - v}  - \frac{1}{2} \right) \textrm{Ei}(-u^2) + e^{- u^2 } \right] , 
\end{align} 
\end{widetext}
the first three terms giving the classical dispersion relation for an
ion-acoustic wave (Ei denotes the exponential integral). Solving these
integrals numerically gives the approximate dispersion relation 
\begin{equation*}
0 \approx 1 + \frac{\omega _{pe}^{2}}{k^{2}v_{Te}^{2}} \left( 1 + \frac{ 2i
\gamma_{\text{cl}} }{k c_s} \right) - \frac{\omega _{pI}^{2}}{\omega ^{2}} - 
\frac{\hbar ^{2}\omega _{pe}^{4}}{m^{2}k^{2}v_{Te}^{6}} \left(0.8 + 0.05i
\right) .
\end{equation*}%
which in the quasi-neutral limit $\omega _{pe}^{2}\gg k^{2}v_{Te}^{2}$ can
be written 
\begin{equation}
\omega = kc_{s}\left( 1+ 0.8 \frac{\hbar ^{2}\omega _{pe}^{2}}{%
m^{2}v_{Te}^{4}}\right) -i\gamma _{\mathrm{cl}}\left( 1 - 3 \frac{\hbar
^{2}\omega _{pe}^{2}}{m^{2}v_{Te}^{4}}\right)  \label{Simplified}
\end{equation}%
where $c_{s}=\left( m_{e}/m_{i}\right) ^{1/2}v_{Te}$ is the classical
ion-acoustic velocity and we have introduced the classical electron Landau
damping, $\gamma_{\text{cl}} = k c_s \sqrt{\pi/8} \sqrt{m_e/m_i}$, in the
cold ion limit \cite{boyd}.

\section{Applicability of model and results}

In deriving our model Eq.\ \eqref{23} and the result \eqref{Simplified} a
series of approximations have been made and we here give a brief
recapitulation of these together with a discussion of their implications for
the applicability of the results.

In order to obtain a closure relation for the BBGKY-hierarchy we have
neglected particle-particle correlations in Eq.\ \eqref{correlations}, which
means disregarding the collisional influence. This approximation is known to
be particle and energy conserving \cite{bonitz}. A general motivation for
neglecting the collisions in our calculations is that the effect of these
have been much studied, see e.g. \cite{spitzer1,spitzer2,Lenard,
Balescu,epperlein1,epperlein2}. Furthermore, since we are utilizing a
perturbative method, the two effects can be studied separately and added
together afterwards provided the collisions also are weak. The relative
magnitude of the collisional influence scales as (see e.g. \cite{spitzer1}) $%
(E_{p}/E_{k})^{3/2}$, where $E_{p}=q^{2}n_{0}^{1/3}/\epsilon _{0}$ is the
characteristic potential energy between nearest neighbors and $%
E_{k}=k_{B}T_{e}=m_{e}v_{Te}^{2}/2$ gives the average kinetic energy of an
electron. To check that the magnitude of exchange corrections is not
necessarily negligible compared to collisions we rewrite the parameter of
the previous section as $(\hbar \omega _{pe}/mv_{Te}^{2})^{2}\sim
(E_{p}/E_{k})(E_{F}/E_{k})$ \cite{manfredi2006}, where $E_{F}$ is the Fermi
energy. Thus we find that the ratio $R$ of \ exchange effects over
collisions scales as 
\begin{equation}
R\sim \frac{E_{F}}{E_{k}^{1/2}E_{p}^{1/2}}.
\end{equation}%
From this we find that we may indeed find a regime where exchange effects
dominate over collisions by choosing a sufficiently high density so that $%
E_{F}\gg E_{p}$ and then keep the temperature relatively modest such that $%
E_{k}$ is not too much larger than $E_{F}$.

In our calculations we furthermore used a Maxwell-Boltzmann background
distribution function instead of a Fermi-Dirac distribution. This was done
mainly due to technical reasons, since it facilitates the solutions of some
of the integrals, but is a good approximation as long as $E_{F}/E_{k}$ is
small. We have also used the long scale length limit, which should be valid
given that the de Broglie wavelength of the particles is short enough
compare to the scale lengths we are interested in.

\section{Discussion}

\label{discussion}

From Eq.\ \eqref{Simplified} we note that the effective ion-acoustic
velocity is \textit{increased}, whereas the damping due to wave-particle
interaction is \textit{decreased} due to the exchange effect. As seen from (%
\ref{Simplified}) the relative magnitude of both these effects is of the
order $H^{2}$, where $H=\hbar \omega _{pe}/mv_{Te}^{2}$. As is shown in,
e.g., Ref.\ \cite{manfredi2006}, plotting the line $H=1$ in a log-log
density temperature diagram divides the parameter space in a classical
regime ($H\ll 1$) and a strong quantum regime ($H\gtrsim 1$). However, such
plots are typically performed in order to illustrate the relative importance
of particle dispersive effects. Within a kinetic formalism particle
dispersion is described by the terms with higher order momentum and spatial
derivatives in the Wigner equation \cite{manfredi2006}. For such terms to be
important, in addition to the parameter $H$ not being too small we also
require the scale lengths under study to be short. Specifically we need the
scale lengths to approach the thermal de Broglie wavelength or shorter. Thus
if we exclude the regime of short scale lengths, as we have done here, the
quantum effect of particle dispersion is guaranteed to be of little
significance. By contrast, we see that exchange effects may very well affect
the long scale behavior of the low-frequency density dynamics. Of particular
interest is the change in the damping term. By approaching the regime $H\sim
1$, Eq. (\ref{Simplified}) suggests that we may more or less completely
suppress Landau damping of ion-acoustic waves. Physically this makes sense,
as classically the particles that are resonantly accelerated for a long time
are rather well localized in phase space, which is then counteracted by the
exchange terms. Strictly the regime $H\sim 1$ does not fit into the
perturbation scheme that we have applied here, but qualitatively we still
expect this result to be valid.

Plasmas where exchange effects can be important occur in e.g. laser-plasma
interaction experiments on solid targets, such as in inertial confinement
fusion schemes. After the compression phase, but before the main part of the
heating has occurred we may have a plasma density of the order $10^{32}%
\mathrm{m}^{-3}$and a temperature $T=4\times 10^{6}\mathrm{K}$, in which
case the Debye length and the de Broglie length are comparable and we have $%
\hbar \omega _{p}/k_{B}T\simeq 1$. For these parameters the plasma is
partially degenerate, and the Fermi temperature $T_{F}$ and the
thermodynamic temperature are comparable. Since we have considered the case $%
\hbar \omega _{p}/k_{B}T\ll 1$ however, a substantially higher value of the
electron temperature and the ordering $T_{F}\ll T$ is consistent with our
calculations.

An important result from this study is the general expression for the
exchange term, as given by Eq. (\ref{23}). This term can describe exchange
modification of any type of processes, e.g. altering the coefficients for
three-wave interaction \cite{Three-wave,Three-wave 2}, adjusting the
Zakharov equations \cite{Zakharov-1972,Karpman} or modifying nonlinear wave
particle interaction processes \cite{Brodin-1997,Manfredi-1997}. The main
restriction is due to the assumption of electrostatic fields. The complexity
of the exchange interaction term in (\ref{23}) in practice forces one to do
perturbative calculations. Since the present formalism captures the full
effect of a distribution function which may be far from equilibrium, it
provides a valuable opportunity to evaluate approaches that %
\textcolor{black}{relies on other types of approximations.} 
%are less general. 
Specifically, in time-dependent density functional theory (TDDFT) \cite{TDFT}
the properties of the system is derived from the electron density only (or
is at least limited to macroscopic quantities), in which case the full
dependence on the detailed momentum distribution is disregarded. Due to the
complexity of many nonlinear plasma systems, such a drastic simplification
may be needed, but at the same time is it essential that the accuracy of the
approach can be evaluated. Results from DFT calculations have been used to
describe electrostatic waves in plasmas, see e.g. Eq. (6) of Ref. \cite%
{Manfredi-DFT}, where the further approximation of the adiabatic local
density approximation (ALDA) has been used.\ See also Ref. \cite%
{NL-ion-acoustic} where the exchange effects on nonlinear ion-acoustic waves
have been studied. However, a difference with our case is that the Fermi
temperature was assumed to be higher than the plasma temperature in these
papers. In a very rough sense the previous results agree with ours, as the
relative importance of the exchange term scale as $\left( \hbar \omega
_{p}/E_{K}\right) ^{2}$ in both cases, noting that the characteristic
kinetic energy $E_{k}$ is the thermal energy $k_{B}T$ in our case and the
Fermi energy $k_{B}T_{F}$ in the case of Ref. \cite{Manfredi-DFT}. However,
in our case the phase velocity of the ion-acoustic waves is increased due to
the exchange interaction, whereas based on Eq. (6) of Ref. \cite%
{Manfredi-DFT} the phase velocity is decreased. Still the interpretation of
this fact can be debated. One possibility is that the approximation of ALDA
to evaluate the exchange potential is too restrictive to capture the
ion-acoustic dynamics accurately. Another possibility is that the results
are indeed sensitive to the ordering of $T$ and $T_{F}$, such that the sign
of the exchange effect is reversed when the ordering is changed. Regardless
of this, is is clear that DFT calculations in general cannot capture the
effects of wave-particle interaction, which is responsible for the wave
damping in our case.

\acknowledgements
This research was supported by the Swedish Research Council Grants \#
2010-3727 and 2012-5644, and the European Research Council Grants \#
204059-QPQV and "ATOMAG" ERC-2009-AdG-20090325\#247452.

%%%%%%%%%%%%%%%%%%%%%%%%%%%%%%%%%%%%%%%%%%%
\appendix*

\section{Long scale length limit of the exchange term}

The calculations leading from Eq.\ \eqref{14-spin} to \eqref{23} are
somewhat complicated and we here show the steps in more detail. Since the
long scale length limit of the left hand side of \eqref{14-spin} is already
known, see e.g. \cite{manfredi2006}, we focus solely on the exchange term on
the right hand side. The first step is to assume as spin independent
distribution function and thereby solve the spin integrals which are then
trivial. We then expand the distribution function and the potential $V$ to
second order in $\mathbf{y}$ and use the identity $y_{j}\exp (i\mathbf{p}%
\cdot \mathbf{y}/\hbar )=-i\hbar \partial _{p}^{j}\exp (i\mathbf{p}\cdot 
\mathbf{y}/\hbar )$ we get 
\begin{widetext} 
\begin{eqnarray} 
	I &=& - \int \frac{dp' dp'' dy dz}{2 (2 \pi \hbar)^{3} } 
	e^{ i \mathbf p \cdot \mathbf y \hbar } 
	 e^{ - i \mathbf p' \cdot ( \mathbf x + \mathbf y /2 - \mathbf z)/\hbar} 
	 e^{ - i \mathbf p'' \cdot ( \mathbf z - \mathbf x + \mathbf y /2 ) / \hbar}  
	(-i\hbar \overleftarrow \partial_p^i)  [ \partial_x^i V ( \mathbf x - \mathbf z) ]   
\notag \\ &&
	\left[ 
	 	f \left( \frac{\mathbf x + \mathbf z}{2}, \mathbf p' \right) 
		f \left( \frac{\mathbf x + \mathbf z}{2} , \mathbf p'' \right) 
		- \frac{ i \hbar}{4} \overleftarrow \partial_p^j 
		f \left( \frac{\mathbf x + \mathbf z}{2} , \mathbf p'' \right) 
		\partial_{\frac{x+z}{2}}^j f \left( \frac{\mathbf x + \mathbf z}{2}, \mathbf p' \right) 
		+ \frac{ i \hbar}{4} \overleftarrow \partial_p^j 
		 f \left( \frac{\mathbf x + \mathbf z}{2}, \mathbf p' \right) 
		\partial_{\frac{x+z}{2}}^j  
		f \left( \frac{\mathbf x + \mathbf z}{2} , \mathbf p'' \right) 
	 \right] , 
\notag \\ &&
\end{eqnarray} 
%\end{widetext}
where the arrow above the derivative signifies the direction in which the
derivative acts. The next step is to perform the integration over $\mathbf{y}
$ which is now straight forward as we only have %\begin{widetext} 
\begin{eqnarray}
&&i\hbar \partial _{p}^{i}\int \frac{dp^{\prime }dp^{\prime \prime }dz}{2}%
\delta \left( \mathbf{p}-\frac{\mathbf{p}^{\prime }+\mathbf{p}^{\prime
\prime }}{2}\right) e^{-i(\mathbf{p}^{\prime }-\mathbf{p}^{\prime \prime
})\cdot (\mathbf{x}-\mathbf{z})/\hbar }[\partial _{x}^{i}V(\mathbf{x}-%
\mathbf{z})]  \notag \\
&&\quad \left[ f\left( \frac{\mathbf{x}+\mathbf{z}}{2},\mathbf{p}^{\prime
}\right) f\left( \frac{\mathbf{x}+\mathbf{z}}{2},\mathbf{p}^{\prime \prime
}\right) -\frac{i\hbar }{4}\overleftarrow{\partial }_{p}^{j}f\left( \frac{%
\mathbf{x}+\mathbf{z}}{2},\mathbf{p}^{\prime \prime }\right) \partial _{%
\frac{x+z}{2}}^{j}f\left( \frac{\mathbf{x}+\mathbf{z}}{2},\mathbf{p}^{\prime
}\right) +\frac{i\hbar }{4}\overleftarrow{\partial }_{p}^{j}f\left( \frac{%
\mathbf{x}+\mathbf{z}}{2},\mathbf{p}^{\prime }\right) \partial _{\frac{x+z}{2%
}}^{j}f\left( \frac{\mathbf{x}+\mathbf{z}}{2},\mathbf{p}^{\prime \prime
}\right) \right]  \notag \\
&&
\end{eqnarray}%
In order to keep there result symmetric more symmetric we make the
substitution 
\begin{align}
\mathbf{p}_{1}& =\mathbf{p}^{\prime }-\mathbf{p}^{\prime \prime } \\
\mathbf{p}_{2}& =\frac{\mathbf{p}^{\prime }+\mathbf{p}^{\prime \prime }}{2}
\end{align}%
and then perform the integration over $\mathbf{p}_{2}$ which is easy due to
the delta function. The result is 
\begin{eqnarray}
&&i\hbar \partial _{p}^{i}\int \frac{dp_{1}dz}{2}e^{-i\mathbf{p}_{1}\cdot (%
\mathbf{x}-\mathbf{z})/\hbar }[\partial _{x}^{i}V(\mathbf{x}-\mathbf{z})] 
\notag \\
&&\quad \left\{ f\left( \frac{\mathbf{x}+\mathbf{z}}{2},\mathbf{p}+\frac{%
\mathbf{p}_{1}}{2}\right) f\left( \frac{\mathbf{x}+\mathbf{z}}{2},\mathbf{p}-%
\frac{\mathbf{p}_{1}}{2}\right) \right.  \notag \\
&&\quad \quad \left. -\frac{i\hbar }{4}\partial _{p}^{j}\left[ f\left( \frac{%
\mathbf{x}+\mathbf{z}}{2},\mathbf{p}-\frac{\mathbf{p}_{1}}{2}\right)
\partial _{\frac{x+z}{2}}^{j}f\left( \frac{\mathbf{x}+\mathbf{z}}{2},\mathbf{%
p}+\frac{\mathbf{p}_{1}}{2}\right) -f\left( \frac{\mathbf{x}+\mathbf{z}}{2},%
\mathbf{p}+\frac{\mathbf{p}_{1}}{2}\right) \partial _{\frac{x+z}{2}%
}^{j}f\left( \frac{\mathbf{x}+\mathbf{z}}{2},\mathbf{p}-\frac{\mathbf{p}_{1}%
}{2}\right) \right] \right\} ,  \notag \\
&&
\end{eqnarray}%
where we have also factored out the momentum derivative on the last two
terms. Finally we make the variable substitution $\mathbf{z}\rightarrow 
\mathbf{x}-\mathbf{z}$ and obtain the required result: 
\begin{eqnarray}
&=&\frac{i\hbar }{2}\partial _{p}^{i}\int dp_{1}dze^{-i\mathbf{p}_{1}\cdot 
\mathbf{z}/\hbar }[\partial _{z}^{i}V(\mathbf{z})]f\left( \mathbf{x}-\frac{%
\mathbf{z}}{2},\mathbf{p}+\frac{\mathbf{p}_{1}}{2}\right) f\left( \mathbf{x}-%
\frac{\mathbf{z}}{2},\mathbf{p}-\frac{\mathbf{p}_{1}}{2}\right)  \notag \\
&&-\frac{\hbar ^{2}}{8}\partial _{p}^{i}\partial _{p}^{j}\int dp_{1}dze^{-i%
\mathbf{p}_{1}\cdot \mathbf{z}/\hbar }[\partial _{z}^{i}V(\mathbf{z})]\left[
f\left( \mathbf{x}-\frac{\mathbf{z}}{2},\mathbf{p}-\frac{\mathbf{p}_{1}}{2}%
\right) \partial _{x}^{j}f\left( \mathbf{x}-\frac{\mathbf{z}}{2},\mathbf{p}+%
\frac{\mathbf{p}_{1}}{2}\right) -f\left( \mathbf{x}-\frac{\mathbf{z}}{2},%
\mathbf{p}+\frac{\mathbf{p}_{1}}{2}\right) \partial _{x}^{j}f\left( \mathbf{x%
}-\frac{\mathbf{z}}{2},\mathbf{p}-\frac{\mathbf{p}_{1}}{2}\right) \right] 
\notag \\
&&
\end{eqnarray}%
\end{widetext}


\begin{thebibliography}{99}
\bibitem{maxwell-note} For dense systems, where exchange effects can be of
importance, the ratio of the Fermi temperature and the thermodynamic
temperature can vary considerably, including both the fully degenerate and
the completely non-degenerate plasma. Here we focus on the latter case,
corresponding to, for example, the later stages in the inertial confinement
fusion regime, not too far from ignition. Furthermore, expanding the
Fermi-Dirac distribution in the classical limit (i.e. when the chemical
potential $\mu \rightarrow -\infty$) gives a correction proportional to $%
\hbar^3 n / m^3 v_T^3 \ll \Lambda_{\text{dB}}^3 / L^3 \ll 1$.
\end{thebibliography}
\end{document}